# A TECHNIQUE FOR DETERMINING THE EXTRAGALACTIC DISTANCE SCALE

Masashi Chiba
Astronomical Institute, Tohoku University, Sendai 980-77, Japan
and
Yuzuru Yoshii
National Astronomical Observatory, Mitaka, Tokyo 181, Japan

## ABSTRACT

We propose a method of distance determination based on the internal structure and dynamics of disk galaxies. The method relies on the universal luminosity profile of a stellar disk represented by an exponential law. Calibrating nearby galaxies with known distances, it is found that the scale length of the disk is tightly correlated with the specific combination of central surface brightness *and* rotational velocity at a characteristic radius of 2.15 scale lengths from the center. This suggests that the scale length of the disk may be used as an indicator for extragalactic distance scale. The application of this relation to M51 and M100 allows us to arrive at the distances of about 6 Mpc and 14 Mpc, respectively, implying a Hubble constant of $H_0 = 92 \sim 94$ km s$^{-1}$ Mpc$^{-1}$.

## 1. INTRODUCTION

Accurate measurement of the distances to galaxies is fundamental for understanding how fast the universe expands as well as how much the universe is aged. For distant galaxies beyond a few megaparsecs, the reliable distance estimates from Cepheids are unavailable, and it is customary to construct a cosmic distance ladder using distance-free empirical correlations among the structural and photometric properties of galaxies (e.g. Tully & Fisher 1977; Faber & Jackson 1976; Tonry & Schneider 1988; Pierce & Tully 1988; van den Bergh 1992).

Most of empirical correlations for galaxies originate from the virial theorem which yields a three-parameter relation of $(luminosity) \propto (velocity)^2 \times (length)$ if the mass-to-light ratio is constant. For some particular systems like spiral galaxies, the virial theorem possibly degenerates into two-parameter relations, because the central surface brightness of stellar disk is known to be in a narrow range, yielding an additional constraint $(luminosity)/(length)^2 \approx$ constant (Freeman 1970). The above dimensional argument explains the background for any sought-after relation for galaxies. It is the tightness of the relation that provides a reliable distance indicator, and the tightness itself depends upon how well the correlated quantities can be defined among many different forms which have the same dimensionality.

A widely used method of measuring the distances to spiral galaxies by Tully & Fisher (1977) is based on either the luminosity - velocity relation or the velocity - diameter relation (the former is more popular and is called the Tully-Fisher relation). More specifically, Tully & Fisher used the total luminosities, HI velocity widths, and apparent diameters. We note that the total luminosity represents the luminous matter in both the bulge and disk. On the other hand, the velocity width, which corresponds to the maximum rotational velocity at larger radii, is more closely related to a dark matter in halo and can be affected by distortion due to a companion galaxy. Thus, the correlation was made between the quantities which represent the different components. When the diameter of a galaxy at a fixed isophote is used as a dimension of length, brighter galaxies are biased in favor of larger intrinsic diameters. All these factors increase the dispersion of the Tully-Fisher relation, which enhances errors in the distances to be estimated.

One of the conditions for establishing an ideal distance indicator is that the correlated quantities are chosen on a well-defined physical basis, so that some uncertainties associated with the correlation can be controlled (Aaronson & Mould 1986). We aim to find a more precise distance indicator by using a sensible combination of observable quantities which represent only the disk with little contamination from the bulge and halo.

In §2 we describe the basic scheme of our method. In §3, using the observed quantities for local calibrator galaxies, we derive the fundamental relation which can be used as a distance indicator. The advantages of our method are discussed in §4. Applying our method to the galaxies of M51 and M100, we estimate their distances and a corresponding Hubble constant in §5.

## 2. METHOD

The radial luminosity profile of a galactic disk is well fitted by an exponential function, $I(r) = I_0 \exp(-r/h)$, where $I_0$ is the extrapolated central surface brightness and $h$ the scale length of the disk (Freeman 1970). Under the assumptions that the disk is in centrifugal equilibrium and the mass-to-light ratio $(M/L)_D$ of the disk is constant, such an exponential profile gives a rotation curve that attains a maximum $V_d/\sqrt{2\pi G h I_0 (M/L)_D} = 0.62 (= f)$ at $r/h = 2.15$. In an actual galaxy, the overall shape of its rotation curve would differ from what is expected solely from the exponential disk unless a bulge and/or dark halo constitutes a minor component. However, such deviation would be rather small at $r/h = 2.15$ where $f$ is close to 0.62, because in most cases this radius is well outside the bulge and well inside the dark halo. Thus, the rotational velocity observed at $r/h = 2.15$ manifests the dynamical state of a luminous matter of the disk.

Given the *distance-independent* observables $I_0$ and $V_d$ for a galaxy, we can derive the intrinsic scale length $h$ using the formula which should hold for the disk:

$$h = \left(\frac{1}{2\pi G f^2 (M/L)_D}\right)\frac{V_d^2}{I_0} \equiv s(M,L)\frac{V_d^2}{I_0} \quad . \tag{1}$$

Then, comparing this intrinsic length $h$ to the observed length $h$ measured in angular units, it is straightforward to determine the distance to the galaxy under consideration.

In the above formula, the quantity $s(M,L)$ contains the mass-to-light ratio $M/L \propto L^x$ which would vary from galaxy to galaxy with some power of its luminosity (e.g., Salucci, Ashman, & Persic 1991). It is difficult to predict the explicit form of $s(M,L)$ from purely theoretical grounds: we shall assume that $s$ is represented as a functional of the distance-independent variable $V_d^2/I_0$. In the next section we constrain a possible form of $s$ using the local calibrator galaxies whose distances have been estimated from Cepheids and other methods.

## 3. LOCAL DISTANCE CALIBRATORS

Table 1 tabulates the adopted data for the local calibrator galaxies. The distance moduli (DM) (Col.(3)) to the galaxies, which have been estimated directly from Cepheids, are taken from de Vaucouleurs (1993). For other galaxies the DMs are the average of a number of independent DM determinations. DMs based on the brightest stars are taken from Tammann (1987). The DM of NGC55 is from the indirect distance estimate using carbon stars (van den Bergh 1992), and the DM of NGC5585 is assumed to be equal to that of M101 because they belong to the same group. For M81, we adopt the long-distance scale (Tammann 1987), but this does not change the following result significantly (see below). Columns 4 and 5 contain the disk scale length $h$ (arcmin) and the central surface brightness $\mu_0$ (mag arcsec$^{-2}$) in the $B$ band, together with their $1\sigma$ errors. Since the decomposition of luminosity profile into the exponential disk includes some uncertainty, we have re-determined $h$ and $\mu_0$ as well as their errors. The references given in Col.(6) originally published the luminosity profile. Presented in Cols.(4)-(5) are the average values of our own estimates with those from the references. Column 7 presents the observed line-of-sight velocity at $r = 2.15h$, while Column 8 for the inclination $i$ which has been evaluated by analyzing the velocity field as given in the references (Col.(9)).

The face-on central brightness $I_0$ ($L_{B\odot}$ pc$^{-2}$) is derived from $\mu_0$, after correction for the absorption and inclination. The absolute $B$-magnitude of the Sun is adopted as $M_{B\odot} = 5.48$. The corrections for the Galactic and internal absorption are taken from *Third Reference Catalogue of Bright Galaxies* (RC3, de Vaucouleurs et al. 1991). We have also derived the inclination of the optical disk from the standard formula, $\cos^2 i = (q^2 - q_0^2)/(1 - q_0^2)$, where $q$ is an axial ratio at an isophote of 25 mag arcsec$^{-2}$ (RC3) and $q_0$ is fixed to be 0.2. This photometry-based inclination is not necessarily equal to the velocity-based inclination given in Table 1. To keep consistency in the analysis below, we use the photometry-based inclination in correcting both $V_d$ and $I_0$.

Figure 1 shows the data in the $(\log h, \log V_d/I_0^{1/2})$ diagram. Filled circles represent the galaxies for which the distances are estimated directly by using Cepheids, while empty circles for those estimated by other means. The tight correlation between these two quantities is apparent, and this correlation is fitted by a linear relation:

$$\log h = a \log V_d/I_0^{1/2} + b \quad , \tag{2}$$

where $a$ and $b$ are constants to be determined by the least-squares fit to the data points. We use only the data with direct distance estimates (filled circles) and weight them by an inverse of the sum of errors in $\log h$ and $\log V_d/I_0^{1/2}$. The best fit line is shown by solid line in Figure 2. The slope of this line is $a = 1.045 \pm 0.072$ and its intercept is $b = -0.658 \pm 0.015$. We call this a standard Case A in Table 2.

There is a freedom for choice of some controversial distance estimates (short or long scale) as well as a freedom for use of an alternative scheme of absorption correction based on *A Revised Shapley-Ames Catalog of Bright Galaxies* (RSA, Sandage & Tammann 1981). In order to examine the sensitivity of the least-squares fit to these modifications, we consider five different Cases B to F and their results are tabulated in Table 2. The label "T" in the column DM of Table 2 stands for exclusive use of Tammann's (1987) distance estimates except for NGC3109 (Sandage & Carlson 1988; Capaccioli et al. 1992) and NGC55 (van den Bergh 1992). The label "vdB" is for van den Bergh's (1992) distance estimates except for NGC247, NGC253 and NGC7793 (Tammann 1987) and NGC3109 (Sandage & Carlson 1988; Capaccioli et al. 1992). When compared to Case A, the most noticeable change appears in the scale lengths of M81 and M101, as shown by triangles in the inset of Figure 1.

The values of $a$ and $b$ in Cases B to F are well within $1\sigma$ error of the corresponding values in Case A. The reason is that a less weight is assigned to these big galaxies because visible fine structures such as spiral arms affect reliable determination of the background exponential disk. In other words, the galaxies with regular luminosity profiles are obviously given a high weight in the least-squares analysis. We further note that the fit is also insensitive to the absorption-correction scheme, either RC3 or RSA.

There should be another correlation between $h$ and $V_d$ if $I_0$ resides in a narrow range and is approximated as a constant. We have confirmed this correlation but with a much larger dispersion. As shown in Figure 1, the fact that inclusion of $I_0$ as a third quantity greatly improves the correlation justifies the specific way of combining $I_0$ with $V_d$ in Eq.(1).

## 4. DISCUSSION

### 4.1. *Inclination Correction*

One of the remarkable points is that the nearly face-on galaxy M101 with $i = 27^\circ$ is close to the line of the best fit in Figure 1, which suggests that our method is also applicable to galaxies with such small inclinations. This is contrasted with the Tully-Fisher relation for which the sample is restricted to highly-inclined galaxies with $i \geq 45^\circ$ to minimize the line-of-sight velocity correction, but with $i < 85^\circ$ to avoid large correction for internal absorption.

Our method uses the relation of the inclination-independent $h$ against the variable $V_d/I_0^{1/2}$ which depends on $i$ as follows:

$$\log \frac{V_d}{I_0^{1/2}} = \log \left( \frac{V_d}{I_0^{1/2}} \right)^{uncor} - \frac{1}{2} \log\{(\sin i)^2 (\cos i)^{1-0.4\alpha(T)}\} \quad , \tag{3}$$

where the superscript *uncorr* denotes the quantity without inclination correction. The factor $\alpha(T)$ stems from the internal absorption, and it depends on galaxy type index $T$. The explicit form of $\alpha(T)$ is given in RC3. Equation (3) is used to figure out how the error in $i$ propagates to $\log V_d/I_0^{1/2}$. Given $\Delta i = 5^\circ$, for example, and $\alpha(T) = 1.4$ for an average type index $T = 6.5$ of local calibrators, we calculate the propagated error $\Delta \log V_d/I_0^{1/2}$ and the result is shown by the solid line in Figure 2.

The above exercise shows that $\Delta \log V_d/I_0^{1/2}$ for M101 ($i \sim 27^\circ$) is as small as $\pm 0.06$ and it is equal to that for highly-inclined galaxies with $i \sim 85^\circ$. Remembering that the slope in Eq.(2) is $a \sim 1$ as shown in Table 2, we can evaluate the error in the resulting determination of a Hubble constant $H_0$ using $\Delta \log V_d/I_0^{1/2} \sim \Delta \log h \sim \Delta \log H_0$. When the sample is restricted to $45^\circ < i < 80^\circ$ as in the Tully-Fisher method, it turns out that the uncertainty of $\log H_0$ due to the inclination correction is well within $\pm 0.02$ and is practically negligible.

### 4.2. *The Mass-to-Light Ratio of the Disk*

The tight relation (Eq.(2)) which we found, guided by a theoretical argument, relies heavily on the local calibrator galaxies. Therefore, the derived relation still needs a physical explanation. The dynamical equation in Eq.(1) is exact for a disk. The scaling relation $h \propto (V_d/I_0^{1/2})^a$ (Eq.(2)) should be equivalent to $f^2(M/L)_D \propto (V_d/I_0^{1/2})^{(2-a)}$. It is this regularity of the internal structure and dynamics of the disk that exists behind the tightness of the derived relation. Our result of $a \sim 1$ as well as $f \sim 0.6$ indicates that the mass-to-light ratio of the disk varies from galaxy to galaxy as $(M/L)_D \propto V_d/I_0^{1/2} \propto (L_D/I_0)^{1/2}$. It is interesting that the value of $a$ is almost exactly equal to an integer of unity (see Table 2). If confirmed from more extensive data, it is tempting to explain its origin within the context of galaxy formation and evolution.

### 4.3. *Summary of the Proposed Method*

Our method uses the three quantities of the rotation velocity, the scale length of the disk, and the central surface brightness of the disk which are designed to represent the disk with little contamination of the bulge and halo.

The rotational velocity $V_d$ at a relatively inner radius of $r/h = 2.15$ is contributed maximally by the exponential mass distribution of the disk. The velocity measured at this radius is almost irrelevant to how much the outer velocity field is distorted due to tidal effects. The scale length $h$ and the central surface brightness $\mu_0$ are determined through the decomposition of the luminosity profile, in a coupled way that the larger $h$ results in fainter $\mu_0$. These quantities are basically independent of the outer cut-off of the disk which results from the isophotal threshold above which the flux is measured.

Local illumination of young stars in spiral arms affects the profile decomposition, but the resulting $h$ and $\mu_0$ change their location, more or less, along the correlation in Fig.1. We see this from the inset where the results of two different decompositions for M31 are shown. It is evident from this figure that the uncertainty of the profile decomposition hardly violates the correlation derived in §3.

On grounds of dimensional argument, our method is transformed to the virial theorem. The application to distance estimates is traced back to Oepik (1922). The present paper basically demonstrates the virial method, but from a practical viewpoint of utilizing modern data to determine distances with precision. The merits of our method are: (1) the result is very insensitive to the inclination correction, so the restriction on $i$ is greatly relaxed, and (2) the correlated quantities have a physical basis and are well defined and unambiguous. The apparent drawback is that detailed observations of photometry and kinematics are required. However, observation with modern equipment can provide the required data for the precise distance measurements advocated here.

## 5. APPLICATION OF THE METHOD

The relation we propose to determine the distance is (Case A):

$$\log h = 1.045(\pm 0.072) \log V_d / I_0^{1/2} - 0.658(\pm 0.015) \quad (\chi^2 = 0.0062) \quad , \qquad (4)$$

where the units are $h$(kpc), $V_d$(km/s), and $I_0(L_{B\odot}\ \mathrm{pc}^{-2})$. We apply this relation to the Whirlpool galaxy, M51, and a member of the Virgo Cluster, M100. The basic quantities of M51 are $h = 111.60$ arcsec, $\mu_0 = 21.56B$ mag arcsec$^{-2}$, $V_d = 164.23$ km s$^{-1}$ and $i = 20°$ (Schweizer 1976; Burkhead 1978; Elmegreen & Elmegreen 1984; Tully 1974), and those for M100 are $h = 56.60$ arcsec, $\mu_0 = 21.33B$ mag arcsec$^{-2}$, $V_d = 213.80$ km s$^{-1}$ and $i = 27°$ (Schweizer 1976; Benedict 1976; Elmegreen & Elmegreen 1984; Kodaira et al. 1986; Distefano et al. 1990). The RC3 scheme has been used for the correction of Galactic and internal absorption, and the inclination has been estimated from the analysis of velocity field of the galaxy. Note that the values of $\mu_0$ and $V_d$ are those corrected for the absorption and inclination.

For M51, the velocity-based inclination $i = 20°$ differs from the photometry-based inclination taken from RC3, most probably due to the strong interaction with the companion galaxy. Since the absorption correction is negligible in a nearly face-on galaxy, we have adopted $i = 20°$, arriving at the distance of $5.97^{+1.17}_{-0.98}$ Mpc. The recession velocity of M51 is adopted as 551 km s$^{-1}$ from RC3, then we obtain the Hubble constant $H_0 = 92.3^{+18.1}_{-15.1}$ km s$^{-1}$ Mpc$^{-1}$.

For M100, the velocity-based inclination $i = 27°$ results in the distance of $14.25^{+1.89}_{-1.67}$ Mpc, while $13.88^{+1.84}_{-1.63}$ Mpc for the photometry-based inclination, thereby arriving at the mean distance of $14.06^{+1.87}_{-1.65}$ Mpc. The systematic recession velocity of the Virgo Cluster is estimated as 1316$\pm$120 km s$^{-1}$, after corrected for the Virgocentric infall taken from Pierce & Tully (1988). With this recession velocity we obtain $H_0 = 93.6^{+15.4}_{-13.2}$ km s$^{-1}$ Mpc$^{-1}$.

We are grateful to the referee for his valuable comments on our method, and to B.A. Peterson and N. Visvanathan for comments on the manuscript.

## REFERENCES

Aaronson, M. & Mould, J. 1986, ApJ, 303, 1

Benedict, G.F. 1976, AJ, 81, 89

Burkhead, M.S. 1978, ApJS, 38, 147

Capaccioli, M., Caon, N., D'Onofrio, M. & Trevisani, S. 1992, AJ, 103, 1151

Carignan, C. 1985a, ApJ, 299, 59

—. 1985b, ApJS, 58, 107

Carignan, C. & Puche, D. 1990a, AJ, 100, 394

—. 1990b, AJ, 100, 641

Comte, G., Monnet, G. & Rosado, M. 1979, A&A, 72, 73

Coté, S., Carignan, C. & Sancisi, R. 1991, AJ, 102, 904

de Vaucouleurs, G. 1958, ApJ, 128, 465

—. 1959, ApJ, 130, 728

—. 1993, ApJ, 415, 10

de Vaucouleurs, G., de Vaucouleurs, A., Corwin, H.G., Jr., Buta, R.J., Paturel, G. & Fouqué, P. 1991, Third Reference Catalogue of Bright Galaxies (New York:Springer-Verlag)

Distefano, A., Rampazzo, R., Chincarini, G. & de Souza, R. 1990, A&AS, 86, 7

Elmegreen, D.M. & Elmegreen, B.G. 1984, ApJS, 54, 127

Elson, R.A.W., Fall, S.M. & Freeman, K.C. 1987, ApJ, 323, 54

Emerson, D.T. 1976, MNRAS, 176, 321

Faber, S.M. & Jackson, R.E. 1976, ApJ, 204, 668

Faber, S.M., et al. 1987, in Nearly Normal Galaxies, ed. Faber, S.M. (New York:Springer-Verlag), 175

Freeman, K.C. 1970, ApJ, 160, 811

Kent, S.M. 1987, AJ, 93, 816

Kodaira, K., Watanabe, M. & Okamura, S. 1986, ApJS, 62, 703

Oepik, E. 1922, ApJ, 55, 406

Okamura, S., Kanazawa, T. & Kodaira, K. 1976, PASJ, 28, 329

Okamura, S., Takase, B. & Kodaira, K. 1977, PASJ, 29, 567

Pence, W.D. 1980, ApJ, 239, 54

Pierce, M.J. & Tully, R.B. 1988, ApJ, 330, 579

Puche, D. & Carignan, C. 1991, ApJ, 378, 487

Puche, D., Carignan, C. & Gorkom, J.H. 1991a, AJ, 101, 456

Puche, D., Carignan, C. & Wainscoat, R.J. 1991b, AJ, 101, 447

Rogstad, D.H., Wright, M.C.H. & Lockhart, I.A. 1976, ApJ, 204, 703

Rogstad, D.H., Crutcher, R.M. & Chu, K. 1979, ApJ, 229, 509

Rots, A.H. 1975, A&A, 45, 43

Salucci, P., Ashman, K.M. & Persic M. 1991, ApJ, 379, 89

Sandage, A. & Tammann, G.A. 1981, A Revised Shapley-Ames Catalog of Bright Galaxies (Washington:Carnegie Institution)

Sandage, A. & Carlson, G. 1988, AJ, 96, 1599

Schweizer, F. 1976, ApJS, 31, 313

Shostak, G.S. 1973, A&A, 24, 411

Tammann, G.A. 1987, in Observational Cosmology, ed. Hewitt, A., Burbidge, G. & Fang, L.Z. (Dordrecht:Reidel), 151

Tonry, J. & Schneider, D.P. 1988, AJ, 96, 807

Tully, R.B. 1974, ApJS, 27, 437

Tully, R.B. & Fisher, J.R. 1977, A&A, 54, 661

van den Bergh, S. 1992, PASP, 104, 861

Visser, H.C.D. 1980, A&A, 88, 149

Yoshizawa, M. & Wakamatsu, K. 1975, A&A, 44, 363

TABLE 1
LOCAL CALIBRATION DATA

| Name | Type | DM (mag) | $\log h$ (arcmin) | $\mu_0$ (mag arcsec$^{-2}$) | Ref. | $\log V_d$ (km s$^{-1}$) | $i$ ($^\circ$) | Ref. |
|---|---|---|---|---|---|---|---|---|
| (1) | (2) | (3) | (4) | (5) | (6) | (7) | (8) | (9) |
| *Local Group* | | | | | | | | |
| LMC[a]............ | SBm | 18.35±0.05 | 1.95±0.03 | 21.61±0.08 | 1 | 1.568±0.001 | 30 | 1 |
| M31[a]............. | Sb | 24.25±0.05 | 1.41±0.03 | 21.20±0.32 | 2,3 | 2.375±0.008 | 78 | 2 |
| M33[a]............. | Sc | 24.43±0.08 | 0.91±0.02 | 21.19±0.19 | 4 | 1.897±0.007 | 55 | 3 |
| NGC3109[a]..... | Sm | 25.70±0.20 | 0.59±0.13 | 21.95±0.14 | 5 | 1.657±0.011 | 80 | 4 |
| *Sculptor Group* | | | | | | | | |
| NGC300[a]....... | Sc | 26.40±0.10 | 0.54±0.05 | 21.86±0.37 | 6 | 1.718±0.022 | 42.3 | 5,6,7 |
| NGC247......... | Sc | 26.70±0.20 | 0.59±0.01 | 22.15±0.11 | 6 | 1.954±0.015 | 74 | 5,6,8 |
| NGC253......... | Sc | 27.50±0.20 | 0.52±0.02 | 20.57±0.18 | 7,9 | 2.245±0.004 | 72 | 6,10 |
| NGC7793....... | Sd | 27.50±0.20 | -0.02±0.01 | 20.19±0.22 | 6 | 1.820±0.005 | 47 | 6,9 |
| NGC55........... | Sc | 25.66±0.13 | 0.55±0.02 | 20.31±0.31 | 8 | 1.817±0.010 | 77 | 6,11 |
| *M81 Group* | | | | | | | | |
| NGC2403[a].... | Sc | 27.30±0.10 | 0.28±0.09 | 20.65±0.16 | 2,10 | 1.998±0.037 | 60 | 12 |
| NGC4236....... | Sd | 27.30±0.10 | 0.45±0.01 | 22.14±0.17 | 2 | 1.893±0.009 | 75 | 12 |
| M81[a]............. | Sb | 28.70±0.20 | 0.44±0.02 | 21.07±0.57 | 2,11,12 | 2.309±0.002 | 59 | 13,14 |
| *M101 Group* | | | | | | | | |
| NGC5585....... | Sd | 28.67±0.15 | -0.10±0.02 | 21.70±0.15 | 13 | 1.761±0.004 | 51.5 | 15 |
| M101[a]........... | Sc | 28.67±0.15 | 0.35±0.08 | 21.50±0.23 | 11,14 | 1.881±0.009 | 27 | 16 |

[a]Cepheid distance estimates are available.

NOTES.—Col.(3):Distance moduli from de Vaucouleurs 1993, except for NGC247, NGC253, NGC7793, M81 (Tammann 1987) and NGC55 (van den Bergh 1992).

Col.(4):Disk scale lengths.

Col.(5):Extrapolated central surface brightnesses of the disks in the $B$ band (uncorrected for inclination and absorption).

Col.(6):References for surface photometry parameters:(1)Elson et al. 1987; (2)Kent 1987; (3)de Vaucouleurs 1958; (4)de Vaucouleurs 1959; (5)Carignan 1985a; (6)Carignan 1985b; (7)Puche et al. 1991a; (8)Puche et al. 1991b; (9)Pence 1980; (10)Okamura et al. 1977; (11)Schweizer 1976; (12)Yoshizawa & Wakamatsu 1975; (13)Coté et al. 1991; (14)Okamura et al. 1976.

Col.(7):Line-of-sight velocities at a radius of $r/h = 2.15$ (uncorrected for inclination).

Col.(8):Inclinations estimated from analysis of velocity fields.

Col.(9):References for rotation velocities:(1)Elson et al. 1987; (2)Emerson 1976; (3)Rogstad et al. 1976; (4)Carignan 1985a; (5)Kent 1987; (6)Puche & Carignan 1991; (7)Rogstad et al. 1979; (8)Carignan & Puche 1990b; (9)Carignan & Puche 1990a; (10)Puche et al. 1991a; (11)Puche et al. 1991b; (12)Shostak 1973; (13)Rots 1975; (14)Visser 1980; (15)Coté et al. 1991; (16)Comte et al. 1979.

TABLE 2
LEAST-SQUARE FITTING PARAMETERS

| Case | DM[a] | Absorption[b] | slope | intercept | $\chi^2$ |
|---|---|---|---|---|---|
| A | deV/T | RC3 | 1.045±0.072 | −0.658±0.015 | 0.0062 |
| B | deV/T | RSA | 1.060±0.142 | −0.668±0.028 | 0.0173 |
| C | T | RC3 | 1.031±0.077 | −0.637±0.017 | 0.0056 |
| D | T | RSA | 1.041±0.136 | −0.641±0.028 | 0.0148 |
| E | vdB | RC3 | 1.011±0.118 | −0.624±0.026 | 0.0089 |
| F | vdB | RSA | 1.022±0.151 | −0.630±0.031 | 0.0018 |

[a]Distance moduli mainly from de Vaucouleurs 1993 (deV), Tammann 1987 (T), and van den Bergh 1992 (vdB). For more explanation see text.
[b]Absorption corrections from RC3 or RSA catalogue.

**Figure Captions**

**FIG. 1** Log-Log plot of the scale length of a disk $h$ (kpc) against the combination of the rotational velocity $V_d$ (km s$^{-1}$) at $r/h = 2.15$ and the extrapolated central surface brightness $I_0$ ($L_{B\odot}$ pc$^{-2}$) for Case A. Filled circles denote the local calibrator galaxies for which the direct distance measurements are available, while empty circles denote the calibrators with only indirect distance measurements. The solid straight line is the result of weighted least-squares fit to the filled circles. *The inset*: Triangles show how the locations of M81 and M101 in the diagram change for Case E with different distance moduli, compared to those in Case A (filled circles). Empty circle for M31 corresponds to one of the different decompositions for its photometric data from which the mean and error are estimated (filled circle). Each line is the result of weighted least-squares fit for Case A (solid), B (dotted), C(short-dashed) and E(long-dashed). Note the very small difference among the fitted lines.

**FIG. 2** The error in $\log V_d/I_0^{1/2}$ arising from the uncertainty of inclination $\Delta i = 5°$. The dependence of internal absorption correction on axial ratio is written as $\alpha \log$(major axis/minor axis) with $\alpha = 1.4$ for the average type index $T = 6.5$ for the local calibrator galaxies in Table 1.

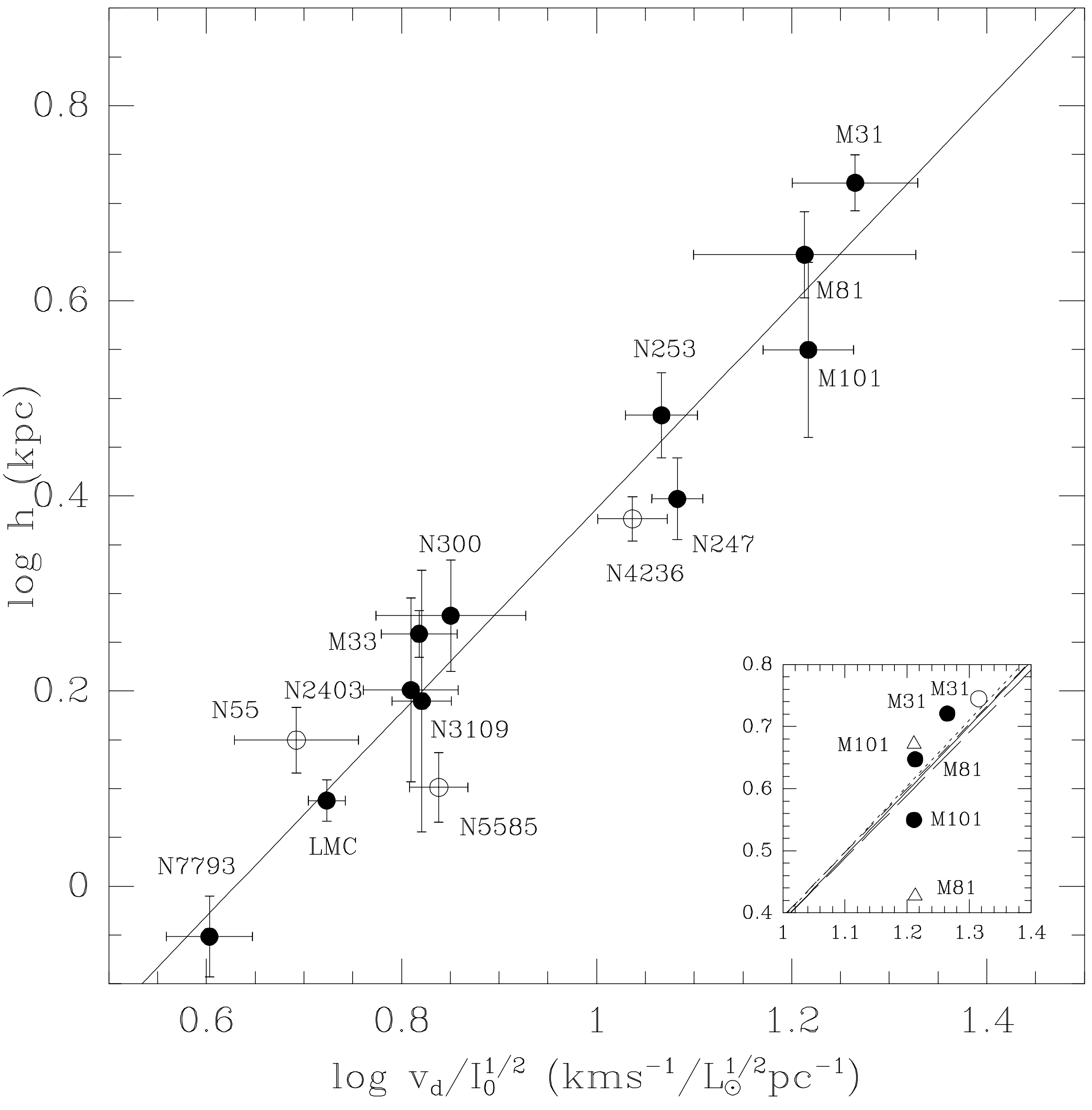

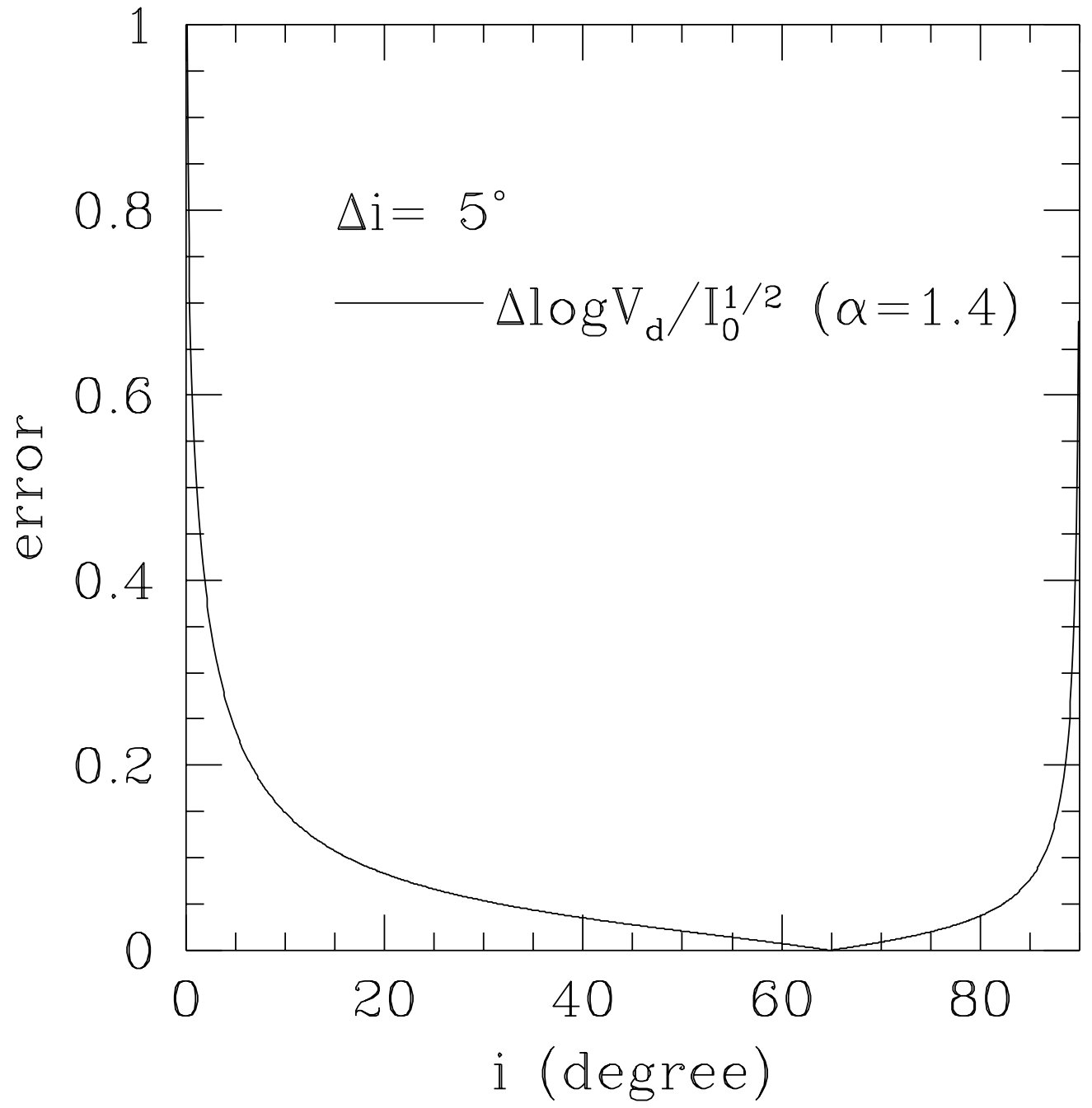